# Optimum signal duration for Human Activity Recognition based on Deep Convolutional Neural Networks

Farhad Nazari[*], Arian Shajari, Darius Nahavandi,
*Member*, and IEEE, Navid Mohajer

*Institute for Intelligent System Research and Innovation (IISRI), Deakin University, Australia*

*Abstract*— Human Activity Recognition (HAR) stands as a pivotal technique within pattern recognition, dedicated to deciphering human movements and actions utilizing one or multiple sensory inputs. Its significance extends across diverse applications, encompassing monitoring, security protocols, and the development of human-in-the-loop technologies. However, prevailing studies in HAR often overlook the integration of human-centered devices, wherein distinct parameters and criteria hold varying degrees of importance compared to other applications. Notably, within this realm, curtailing the sensor observation period assumes paramount importance to safeguard the efficiency of exoskeletons and prostheses. This study embarks on the optimization of this observation period specifically tailored for HAR using Inertial Measurement Unit (IMU) sensors. Employing a Deep Convolutional Neural Network (DCNN), the aim is to identify activities based on segments of IMU signals spanning durations from 0.1 to 4 seconds. Intriguingly, the outcomes spotlight an optimal observation duration of 0.5 seconds, showcasing an impressive classification accuracy of 99.95%. This revelation holds immense significance, elucidating the criticality of precise temporal analysis within HAR, particularly concerning human-centric devices. Such findings not only enhance our understanding of the optimal observation period but also lay the groundwork for refining the performance and efficacy of devices crucially relied upon for aiding human mobility and functionality.

## I. Introduction

Human activity recognition (HAR) is a pattern recognition technique that is utilised to interpret people's movements and actions using different sensing technologies [1-3]. HAR has several applications in the present times [4]. They could be used in the medical, military, security applications, surveillance, gaming, entertainment, sports and a number of other fields [5-8].

An example of a HAR model being used in the medical field is its utilisation in differentiating the movements of patients with early-stage Parkinson's disease and healthy controls [9]. In one study, a HAR model called the SmartARM system was developed to create a comprehensive status of soldiers based on their physical activity levels [10]. In another study, a deep learning technique that can categorise and pinpoint identified activities was trained to specifically recognise typical and atypical activities for security monitoring purposes [11]. An instance of HAR's application in gaming can be the development of the human motion capture and pose recognition model by Huo et al. [12].

Shechtman and Irani [13] were able to identify dives into a pool during a swimming relay match by employing an extended behaviour-based similarity measure, which could effectively differentiate most dives from other activities despite the presence of significant noise and numerous concurrent activities.

There are two general approaches for data collection in HAR: sensor-based and vision-based [14, 15]. The vision-based method of data collection requires a mode of image or video collection followed by activity detection and recognition [16]. For instance, a study developed a machine learning model to recognise certain human activities, such as walking, running and hand waving, using a database [17], which contains image sequences of these actions [18]. As mentioned, the other method of HAR data collection is sensor-based. With the developments in sensor technology in the last decades, including improvements in their computational power, size, accuracy and manufacturing cost, they have become more accessible and widely used [19, 20]. For the purpose of HAR, different types of sensors could be utilised [21]. For example, in one study, radio frequency identification-based sensors were used to detect certain indoor human activities [22]. Components of smartphones, such as accelerometers, have also been used for this purpose [23]. Another instance of sensors being used for HAR is in a study conducted by Jia and Bin [24], where a hybrid of accelerometer and ECG sensors was used for HAR. In another study conducted by Ashry et al. [25], a deep learning algorithm was used to classify human activity using data collected by an Apple watch, which was used as a wearable Inertial Measurement Unit (IMU) sensor.

When wearable sensors are utilised in HAR, different placements of sensors on a human body could be valid, and the accuracy of their results could be impacted by this factor and based on the activities that are being detected. An instance of this is a system developed in a study by Nam and Park [26], in which accelerometer sensors are attached to children's waists to detect certain activities that might put them in danger. While in a study conducted by Hemalatha and Vaidehi [27], placed a device with accelerometer sensors on their experiment participants' chests to detect falling of the elderly. Also, in a study conducted by Mannini et al. [28], a single accelerometer was placed on the wrist or ankle of its participants to develop a system that would recognise certain common and un-common daily activities.

The methods or the models utilised for the detection and recognition of activities are another aspect of HAR [29]. AI based models can be used for this purpose as they have in other advanced fields [30-35]. Based on the nature of the collected data, size of data and other factors, existing studies have utilised a variety of models for this aim. Statistical learning methods, naive Bayes and k-nearest neighbour, were utilised in a study conducted by Chavarriaga et al. [36], to classify certain human activities by data collected by a number of different sensors. A study compared different machine learning classifiers, which were decision tree, multi-layer perception, naive Bayes, logistic, and k-nearest neighbour, to find the most accurate model for HAR, using a dataset gathered from accelerometers and gyroscopes [37]. In another study, conducted by Sharma et al. [38], a neural network classifier was designed to classify human activity. Moreover, a study developed a deep convolutional neural network to perform a HAR using a smartphone sensor [39].

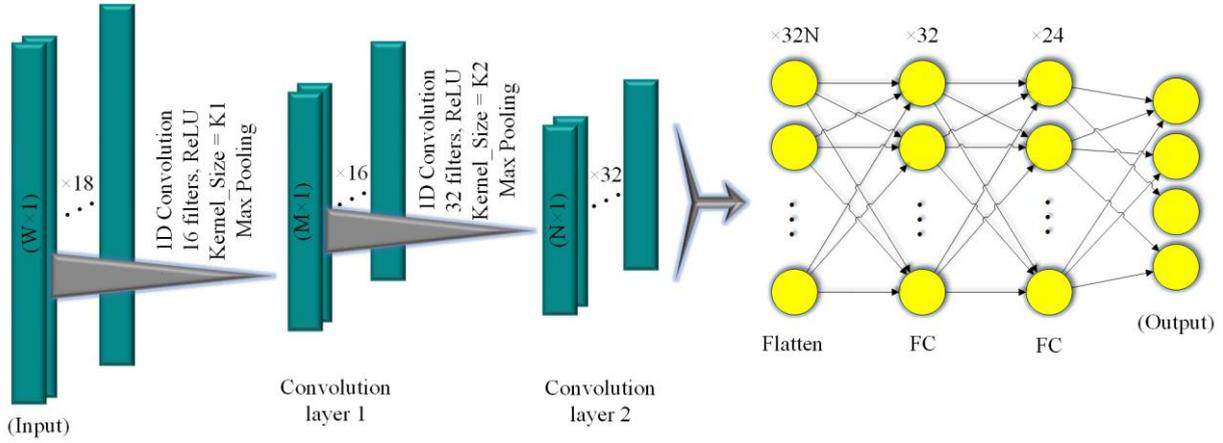

*Fig. 1. The architecture of the proposed CNN model. W is the size of the input layer or the sample length, and M and N are the output size of first and second convolution layers, respectively. M and N depends on W, K1 and K2.*

While the literature is rich with promising studies in HAR, most of them are not directly aimed at human-in-the-loop technologies [40]. In these applications, the importance of parameters and criteria are different from the ones in other applications. While the sensor observation period to make a prediction may not be of great importance in IoT applications, it is crucial to reduce this number as much as possible in exoskeletons and prostheses before hurting the performance for real-time applications. This study aims to find optimum signal observation period for the purpose of HAR from Inertial Measurement Unit (IMU) sensors for use in wearable assistive devices. A deep Convolutional Neural Network (CNN) has been proven to perform exceptionally well in HAR from portions of IMU signals, showing up to 99.98% classification accuracy on the same dataset [41, 42].

The rest of the paper is organised as follows. The next section describes the dataset and our pre-processing methods, followed by the proposed model described in section III. Section IV will contain a discussion of the results, followed by a concluding section V that summarises this research.

## II. DATASET AND PRE-PROCESSING

PAMAP2 Physical Activity Monitoring dataset [43] was used in this study. Nine subjects performed 18 different activities while three IMU sensors collected motion signals from their chest, dominant wrist, and ankle in three dimensions at 100Hz.

In the dataset [44], the duration of participation of the subjects in activities is not consistent; i.e. not all participants performed all activities, and their ratio is inconsistent. What is more, the ability to distinguish between all of these activities will not give us a significant edge in controlling an assistive device. We decided to continue the research with five routine activities: sitting, standing, walking, and ascending and descending stairs.

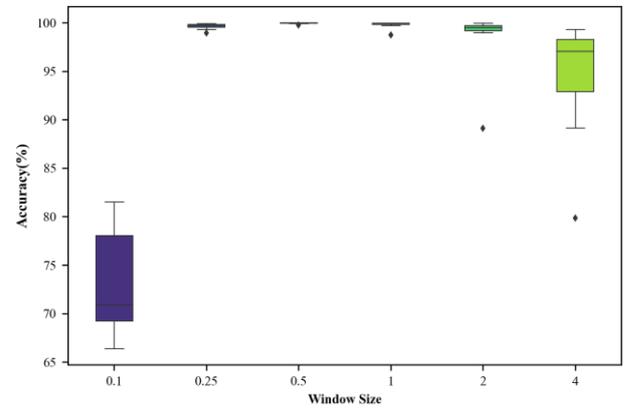

*Fig. 2. Accuracy distribution of the proposed model on different window sizes.*

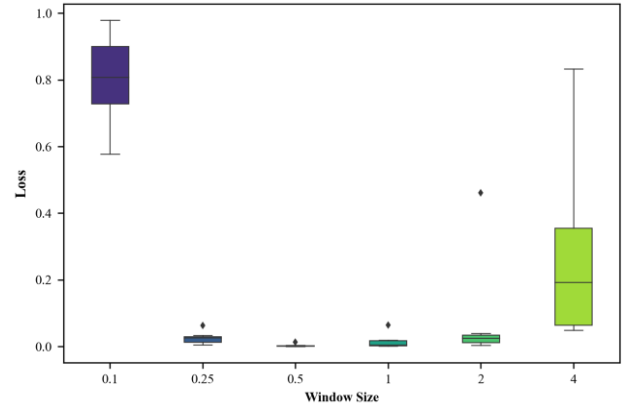

*Fig. 3. Cross-entropy loss value distribution of the model on different window sizes.*

The signals contain 3D angular and linear acceleration data. In other words, it is a multimodal time-series signal with a modality of 18. After normalising, these signals have been transformed to have a mean of zero and a standard deviation of one, then into a tensor with three dimensions, representing window-size or the duration of the signal that the model will get as an input in each state, the modality of the data, and the number of samples.

The variable in this study is the window size to find the optimum duration of observation to detect human activity. This study evaluates the performance of a CNN model on 0.1, 0.25, 1, 2 and 4 seconds observation of the data. We hypothesise that larger window sizes lead to better performance as the signal can carry more complex information. However, for real-time applications, this value must be reduced as much as possible while maintaining prediction accuracy. These window sizes have been chosen to test the limits of our approach and find the minimum acceptable window size. For 3D transformation, each window had a 25% slide, meaning it had a 75% overlap with the previous sample.

## III. MODEL DEVELOPMENT

This research focuses on optimising the signal length of inertial sensors to detect human activity. The proposed model is a Deep Convolutional Neural Network (DCNN) with two 1D convolution layers, followed by another two fully connected layers. Multi-layer convolutional layers perform exceptionally in detecting higher levels of features from low-level signals. Fig 1 shows the proposed model used in this study, with the M and N being the size of the input to the layer as in (1) and (2):

$$M = [(W - K1 + 2P)/S] + 1 \quad (1)$$
$$N = [(M - K2 + 2P)/S] + 1 \quad (2)$$

In these equations, w is the input length, K1 and K2 are the kernel size of convolution layers 1 and 2, padding (P) is zero, and stride (S) is one. We previously showed that this model performs exceptionally well when K1 and K2 are 7 and 11, respectively. However, these numbers were too big for window sizes of 0.1 and 0.25 seconds, returning negative dimensions. For these input sizes, kernel sizes 3 and 5 have been used.

The proposed model uses 16 and 32 filters in the first and second convolution layers and Rectified Linear Unit (ReLU) activation function. The hidden FC layers use 32 and 24 neurons, respectively. Each convolution layer is followed by a max-pooling layer to half their output size while maintaining important information. The output of these layers was passed through two fully connected layers with 32 and 24 neurons with ReLU activation function. The output layer has five neurons and utilises the SoftMax activation function. The model has been compiled with Adam optimisation and a 0.3 dropout rate in each step to prevent overfitting.

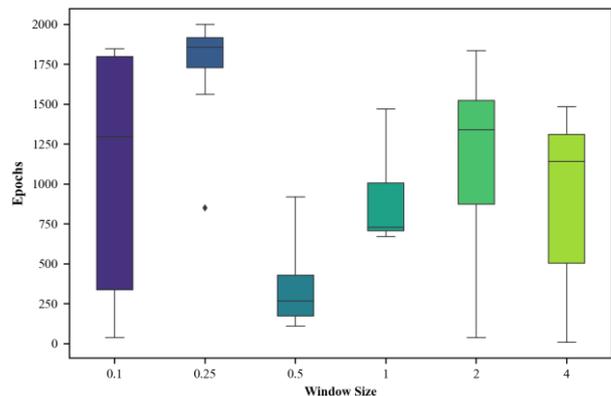
*Fig. 4. The number of epochs required to train the model on different window sizes.*

## IV. RESULTS AND DISCUSSION

In this study, we evaluated the performance of a DCNN model for HAR classification using different observation window sizes of inertial signals. To achieve this, we used an 8-fold cross-validation approach to train the model on inertial data with observation sizes ranging from 0.1 to 4 seconds and then evaluated its performance by measuring its accuracy and categorical cross-entropy loss values.

Table 1 presents the results of the study and displays the k-fold cross-validation accuracy of the proposed model for various window sizes. Our analysis revealed that the model performed best with 0.5-second window size, achieving an average accuracy of 99.95% and a low categorical cross-entropy loss value of 0.003. This suggests that the model was highly effective in identifying patterns in the inertial data, leading to accurate classification results. Furthermore, we found that the model's performance was consistent with this window size, with only a 0.08% standard deviation in accuracy across eight trials. This is an important finding, as it indicates that the model can be used reliably in real-world applications that require precise and consistent results.

We also observed that the CNN model achieved relatively high accuracy with observation periods of 1 and 0.25 seconds, with accuracy values of 99.77% and 99.63%, respectively, and loss values of 0.01 and 0.03. However, these results were not as optimal as the 0.5-second window size. Our analysis suggests that the optimal observation window size for this model is likely to be between 0.25 and 1 second, and perhaps closer to 0.5 second. This could be attributed to the fact that a 0.5-second window strikes a balance between capturing enough data for accurate classification while minimising the model's size and complexity. The slightly lower but still reasonable performance of the model on the 2-second window size is yet another proof on the hypothesis.

TABLE 1. The 8-fold cross validation results of training the proposed models on different observation periods.

| Input Size (sec) | Accuracy (%) | acc_std | Loss | loss_std | Epochs | epochs_std |
|---|---|---|---|---|---|---|
| 0.1 | 73.15 | 5.834 | 0.802 | 0.135 | 1082 | 776 |
| 0.25 | 99.63 | 0.332 | 0.026 | 0.018 | 1719 | 375 |
| 0.5 | 99.95 | 0.083 | 0.003 | 0.004 | 350 | 267 |
| 1 | 99.77 | 0.424 | 0.015 | 0.021 | 878 | 279 |
| 2 | 98.23 | 3.695 | 0.076 | 0.156 | 1175 | 610 |
| 4 | 94.17 | 6.652 | 0.266 | 0.270 | 895 | 588 |

The model's performance suffered significantly when the observation time was either too short or too long. In particular, the 0.1-second window size had the poorest performance with an average accuracy of 73% and a loss value of 0.8, indicating that the information in such a short signal was insufficient for the model to distinguish between activities. Meanwhile, the 4-second window size showed a slightly better performance compared to the 0.1-second window size, but was still inferior to other window sizes, which might be due to increased model size and complexity. The authors hypothesised that since the data size is fixed, increasing the model's parameters would lead to an under-fitted model, resulting in a drop in performance. To assess the model's performance for each window size more thoroughly, Figs 2 and 3 present the accuracy and loss value distributions of the proposed model over eight trials for each window size, providing a more comprehensive understanding of the model's performance across different window sizes.

Fig 4 shows the number of training epochs the model needs to fit for each window size. On average, it requires 350 training rounds to fit the model on a 0.5-second window size. This number and its variance increase with window size. Interestingly, the 0.25-second window size required the highest training epochs to fit the model, indicating that the algorithm has a hard time fitting the model to the data. On the other hand, the model shows the most variance in eight-time training when working with the 0.1-second window, which indicates that this signal duration is too short and may not provide enough information for the model to accurately distinguish between different activities; any random variation in the signal can affect the model training as well as its performance.

## V. STUDY LIMITATIONS AND FUTURE WORK

The main limitation of this study is the limited dataset, which may not be representative of all human activities. Future research can consider exploring a larger and more diverse dataset to enhance the generalizability of the findings. Furthermore, the study only considered the performance of the model in a controlled environment, and it is unclear how the model will perform in real-world scenarios. Future studies can investigate the performance of the model in real-world settings, such as in outdoor activities or in diverse populations with different demographics. In the following, we suggest more areas needed to be further explored for practical and applicable literature in the field of HAR:

- Investigating the impact of sensor placement on activity recognition: The placement of sensors on the body can affect the accuracy of activity recognition. Future research could investigate the optimal sensor placement for different activities and explore how different sensor configurations impact activity recognition accuracy.

- Exploring the use of multimodal data for HAR: Human activity involves a complex combination of movements and actions, and incorporating multiple sources of data such as video, audio, and physiological signals could improve activity recognition accuracy. Future research could explore the use of multimodal data for activity recognition.

- Investigating the transferability of HAR models: Activity recognition models trained on one dataset may not perform well on another dataset. Future research could investigate the transferability of activity recognition models between different datasets and explore techniques for improving transferability.

- Incorporating contextual information into activity recognition models: Activities are often performed in specific contexts, such as a home, office, or gym. Incorporating contextual information such as time of day, location, and social context could improve the accuracy of activity recognition models. Future research could explore the use of contextual information for activity recognition.

## VI. CONCLUSION AND FUTURE DIRECTIONS

This study proposed a methodology to identify the ideal signal observation time for accurately detecting human activity from inertial signals. A deep convolutional neural network model was employed, and it was trained on different window sizes ranging from 0.1 to 4 seconds. The model's performance was then assessed using an 8-fold cross-validation method. The results of the study demonstrate that the 0.5-second window size yielded the best performance, with the highest classification accuracy, lowest categorical cross-entropy loss value, and the least number of training epochs required to fit the model. Furthermore, this observation period led to the most consistent results in eight trials, surpassing other window sizes with minimum standard deviation in all metrics. Therefore, the findings suggest that the 0.5-second window size is the most suitable for accurately detecting human activity from inertial signals, which can be beneficial in various fields, including healthcare and sports science.

REFERENCES

[1] O. Banos, J.-M. Galvez, M. Damas, H. Pomares, and I. J. S. Rojas, "Window size impact in human activity recognition," vol. 14, no. 4, pp. 6474-6499, 2014.

[2] E. Kim, S. Helal, and D. J. I. p. c. Cook, "Human activity recognition and pattern discovery," vol. 9, no. 1, pp. 48-53, 2009.

[3] F. Nazari, D. Nahavandi, N. Mohajer, and A. Khosravi, "Human activity recognition from knee angle using machine learning techniques," in *2021 IEEE International Conference on Systems, Man, and Cybernetics (SMC)*, 2021: IEEE, pp. 295-300.

[4] M. Ziaeefard and R. J. P. R. Bergevin, "Semantic human activity recognition: A literature review," vol. 48, no. 8, pp. 2329-2345, 2015.

[5] P. Woznowski, R. King, W. Harwin, and I. Craddock, "A human activity recognition framework for healthcare applications: ontology, labelling strategies, and best practice," in *International Conference on Internet of Things and Big Data*, 2016, vol. 2: SciTePress, pp. 369-377.

[6] O. D. Lara, M. A. J. I. c. s. Labrador, and tutorials, "A survey on human activity recognition using wearable sensors," vol. 15, no. 3, pp. 1192-1209, 2012.

[7] A. Taha, H. H. Zayed, M. Khalifa, and E.-S. M. El-Horbaty, "Human activity recognition for surveillance applications," in *Proceedings of the 7th International Conference on Information Technology*, 2015, pp. 577-586.

[8] F. Demrozi, G. Pravadelli, A. Bihorac, and P. J. I. A. Rashidi, "Human activity recognition using inertial, physiological and environmental sensors: A comprehensive survey," vol. 8, pp. 210816-210836, 2020.

[9] W.-Y. Cheng *et al.*, "Human activity recognition from sensor-based large-scale continuous monitoring of Parkinson's disease patients," in *2017 IEEE/ACM International Conference on Connected Health: Applications, Systems and Engineering Technologies (CHASE)*, 2017: IEEE, pp. 249-250.

[10] A. Mukherjee, S. Misra, P. Mangrulkar, M. Rajarajan, and Y. Rahulamathavan, "SmartARM: A smartphone-based group activity recognition and monitoring scheme for military applications," in *2017 IEEE International Conference on Advanced Networks and Telecommunications Systems (ANTS)*, 2017: IEEE, pp. 1-6.

[11] A. Sunil, M. H. Sheth, and E. Shreyas, "Usual and unusual human activity recognition in video using deep learning and artificial intelligence for security applications," in *2021 Fourth International Conference on Electrical, Computer and Communication Technologies (ICECCT)*, 2021: IEEE, pp. 1-6.

[12] F. Huo, E. Hendriks, P. Paclik, and A. H. Oomes, "Markerless human motion capture and pose recognition," in *2009 10th Workshop on Image Analysis for Multimedia Interactive Services*, 2009: IEEE, pp. 13-16.

[13] E. Shechtman and M. Irani, "Space-time behavior based correlation," in *2005 IEEE Computer Society Conference on Computer Vision and Pattern Recognition (CVPR'05)*, 2005, vol. 1: IEEE, pp. 405-412.

[14] Z. Hussain, M. Sheng, and W. E. J. a. p. a. Zhang, "Different approaches for human activity recognition: A survey," 2019.

[15] L. Chen, J. Hoey, C. D. Nugent, D. J. Cook, Z. J. I. T. o. S. Yu, Man,, and P. C. Cybernetics, "Sensor-based activity recognition," vol. 42, no. 6, pp. 790-808, 2012.

[16] X. Xu, J. Tang, X. Zhang, X. Liu, H. Zhang, and Y. J. s. Qiu, "Exploring techniques for vision based human activity recognition: Methods, systems, and evaluation," vol. 13, no. 2, pp. 1635-1650, 2013.

[17] C. Schuldt, I. Laptev, and B. Caputo, "Recognizing human actions: a local SVM approach," in *Proceedings of the 17th International Conference on Pattern Recognition, 2004. ICPR 2004.*, 2004, vol. 3: IEEE, pp. 32-36.

[18] S. Danafar and N. Gheissari, "Action recognition for surveillance applications using optic flow and SVM," in *Asian Conference on Computer Vision*, 2007: Springer, pp. 457-466.

[19] L. M. Dang, K. Min, H. Wang, M. J. Piran, C. H. Lee, and H. J. P. R. Moon, "Sensor-based and vision-based human activity recognition: A comprehensive survey," vol. 108, p. 107561, 2020.

[20] Y. Liu, L. Nie, L. Liu, and D. S. Rosenblum, "From action to activity: Sensor-based activity recognition," *Neurocomputing,* vol. 181, pp. 108-115, 2016/03/12/ 2016, doi: https://doi.org/10.1016/j.neucom.2015.08.096.

[21] K. Chen, D. Zhang, L. Yao, B. Guo, Z. Yu, and Y. J. A. C. S. Liu, "Deep learning for sensor-based human activity recognition: Overview, challenges, and opportunities," vol. 54, no. 4, pp. 1-40, 2021.

[22] G. A. Oguntala *et al.*, "SmartWall: novel RFID-enabled ambient human activity recognition using machine learning for unobtrusive health monitoring," vol. 7, pp. 68022-68033, 2019.

[23] J. R. Kwapisz, G. M. Weiss, and S. A. J. A. S. E. N. Moore, "Activity recognition using cell phone accelerometers," vol. 12, no. 2, pp. 74-82, 2011.

[24] R. Jia and B. Liu, "Human daily activity recognition by fusing accelerometer and multi-lead ECG data," in *2013 IEEE International Conference on Signal Processing, Communication and Computing (ICSPCC 2013)*, 2013: IEEE, pp. 1-4.

[25] S. Ashry, R. Elbasiony, and W. Gomaa, "An LSTM-based descriptor for human activities recognition using IMU sensors," in *Proceedings of the 15th International Conference on Informatics in Control, Automation and Robotics, ICINCO*, 2018, vol. 1, pp. 494-501.

[26] Y. Nam and J. W. Park, "Child activity recognition based on cooperative fusion model of a triaxial accelerometer and a barometric pressure sensor," *IEEE journal of biomedical health informatics*
vol. 17, no. 2, pp. 420-426, 2013.

[27] C. S. Hemalatha and V. Vaidehi, "Frequent Bit Pattern Mining Over Tri-axial Accelerometer Data Streams for Recognizing Human Activities and Detecting Fall," *Procedia Computer Science,* vol. 19, pp. 56-63, 2013/01/01/ 2013, doi: https://doi.org/10.1016/j.procs.2013.06.013.

[28] A. Mannini *et al.*, "Activity recognition using a single accelerometer placed at the wrist or ankle," vol. 45, no. 11, p. 2193, 2013.

[29] F. Nazari, N. Mohajer, D. Nahavandi, and A. Khosravi, "Comparison of gait phase detection using traditional machine learning and deep learning techniques," in *2022 IEEE International Conference on Systems, Man, and Cybernetics (SMC)*, 2022: IEEE, pp. 403-408.

[30] A. Shajari *et al.*, "Detection of Driving Distractions and Their Impacts: A Comprehensive Review," *Available at SSRN 4141286.*

[31] M. R. C. Qazani, H. Asadi, A. Shajari, Z. Najdovski, C. P. Lim, and S. Nahavandi, "A Development of Time-Varying Weight Model Predictive Control for Autonomous Vehicles," in *2023 IEEE International Conference on Systems, Man, and Cybernetics (SMC)*, 2023: IEEE, pp. 3480-3486.

[32] A. Shajari, H. Asadi, S. Alsanwy, and S. Nahavandi, "Detection of Driver Cognitive Distraction Using Driver Performance Measures, Eye-Tracking Data and a D-FFNN Model," in *2023 IEEE International Conference on Systems, Man, and Cybernetics (SMC)*, 2023: IEEE, pp. 2093-2099.

[33] S. Alsanwy *et al.*, "Prediction of Vehicle Motion Signals for Motion Simulators Using Long Short-Term Memory Networks," in *2022 IEEE International Conference on Systems, Man, and Cybernetics (SMC)*, 2022: IEEE, pp. 34-39.

[34] M. R. C. Qazani *et al.*, "A Prediction of Time Series Driving Motion Scenarios Using LSTM and ESN," in *2022 IEEE International Conference on Systems, Man, and Cybernetics (SMC)*, 2022: IEEE, pp. 1592-1599.

[35] S. Alsanwy, H. Asadi, M. R. C. Qazani, S. Mohamed, and S. Nahavandi, "A CNN-LSTM Based Model to Predict Trajectory of Human-Driven Vehicle," in *2023 IEEE International Conference on Systems, Man, and Cybernetics (SMC)*, 2023: IEEE, pp. 3097-3103.

[36] R. Chavarriaga *et al.*, "The Opportunity challenge: A benchmark database for on-body sensor-based activity recognition," vol. 34, no. 15, pp. 2033-2042, 2013.

[37] W. Wu, S. Dasgupta, E. E. Ramirez, C. Peterson, and G. J. J. J. o. m. I. r. Norman, "Classification accuracies of physical activities using smartphone motion sensors," vol. 14, no. 5, p. e2208, 2012.

[38] A. Sharma, Y.-D. Lee, and W.-Y. Chung, "High accuracy human activity monitoring using neural network," in *2008 third international conference on convergence and hybrid information technology*, 2008, vol. 1: IEEE, pp. 430-435.

[39] C. A. Ronao and S.-B. J. E. s. w. a. Cho, "Human activity recognition with smartphone sensors using deep learning neural networks," vol. 59, pp. 235-244, 2016.

[40] F. Nazari, N. Mohajer, D. Nahavandi, A. Khosravi, and S. Nahavandi, "Applied Exoskeleton Technology: A Comprehensive Review of Physical and Cognitive Human-Robot Interaction," *IEEE Transactions on Cognitive and Developmental Systems,* 2023.

[41] F. Nazari, N. Mohajer, D. Nahavandi, A. Khosravi, and S. Nahavandi, "Comparison study of inertial sensor signal combination for human activity recognition based on convolutional neural networks," in *2022 15th International Conference on Human System Interaction (HSI)*, 2022: IEEE, pp. 1-6.

[42] F. Nazari, D. Nahavandi, N. Mohajer, and A. Khosravi, "Comparison of deep learning techniques on human activity recognition using ankle inertial signals," in *2022 IEEE International Conference on Systems, Man, and Cybernetics (SMC)*, 2022: IEEE, pp. 2251-2256.

[43] A. Reiss and D. Stricker, "Introducing a new benchmarked dataset for activity monitoring," in *2012 16th international symposium on wearable computers*, 2012: IEEE, pp. 108-109.

[44] A. Reiss and D. Stricker, "Creating and benchmarking a new dataset for physical activity monitoring," in *Proceedings of the 5th International Conference on PErvasive Technologies Related to Assistive Environments*, 2012, pp. 1-8.